\documentstyle[graphicx]{mn}

\title{The complex soft X-ray spectrum of NGC 4151}

\author[N.J.\,Schurch, R.S.\,Warwick, R.E.\,Griffiths \& S.M.\,Kahn]
{N.J.\,Schurch$^{1*}$, R.S.\,Warwick$^{2}$, R.E.\,Griffiths$^{1}$, S.M.\,Kahn$^{3}$\\
$^{1}$Department of Physics, Carnegie Mellon University, 5000 Forbes Avenue, 
Pittsburgh, PA 15213 \\
$^{2}$Department of Physics and Astronomy, University of Leicester, 
University Road, Leicester, LE1 7RH \\
$^{3}$ University of Columbia Astrophysics Laboratory, 550 West 120th Street, New York, 10027 \\
$^{*}$ schurch@andrew.cmu.edu}
\date{}

\def\exo{{\it EXOSAT\/}}
\def\gin{{\it Ginga\/}}
\def\ro{{\it ROSAT\/}}

\def\asca{{\it ASCA\/}}

\def\xte{{\it RXTE\/}}
\def\xmm{{\it XMM-Newton\/}}

\def\cha{{\it Chandra\/}}
\def\bep{{\it BeppoSAX\/}}
\def\int{{\it INTEGRAL\/}}
\def\cgro{{\it CGRO\/}}

\def\H0{{\rm ~km~s^{-1}~Mpc^{-1}}}

\def\arcs{{\hbox{$^{\prime\prime}$}}}

\def\etal{et al.~\/}

\def\la{\mathrel{\hbox{\rlap{\hbox{\lower4pt\hbox{$\sim$}}}{\raise2pt\hbox{$<$}}}}}
\def\ga{\mathrel{\hbox{\rlap{\hbox{\lower4pt\hbox{$\sim$}}}{\raise2pt\hbox{$>$}}}}}
\def\ls{\mathrel{\hbox{\rlap{\hbox{\lower4pt\hbox{$\sim$}}}\hbox{$<$}}}}
\def\gs{\mathrel{\hbox{\rlap{\hbox{\lower4pt\hbox{$\sim$}}}\hbox{$>$}}}}

\def\d25{D$_{\rm 25}$}

\def\.25{0.25 keV\thinspace}

\begin{document}

\maketitle 

\begin{abstract}

We present a detailed analysis of the complex soft X-ray spectrum of NGC 4151 measured by the RGS instruments aboard \xmm. The \xmm ~RGS spectra demonstrate that the soft X-ray emission is extremely rich in X-ray emission lines and radiative recombination continua (RRC), with no clear evidence for any underlying continuum emission. Line emission, and the associated RRC, are clearly detected from hydrogen-like and helium-like ionization states of neon, oxygen, nitrogen and carbon. The measured lines are blueshifted with a velocity of between $\sim$100-1000 km s$^{-1}$, with respect to the systemic velocity of NGC 4151, approximately consistent with the outflow velocities of the absorption lines observed in the UV spectrum of NGC 4151 (Kriss \etal 1995), suggestive of an origin for the UV and soft X-ray emission in the same material. Plasma diagnostics from the observed helium-like triplets, imply a range of electron temperatures of $\sim$1-5$\times10^{4}$ K and electron densities of between 10$^{8}$-10$^{10}$ cm$^{-3}$. The soft X-ray spectrum of NGC 4151 is extremely similar to that of NGC 1068, both in terms of the atomic species present and in terms of the relative strengths of the observed emission lines and RRC (Kinkhabwala \etal 2002), suggesting that the soft X-ray excesses observed in many Seyfert 2 galaxies may be composed of similar emission features. Modelling the RGS spectra in terms of emission from photoionized and photoexcited gas in an ionization cone reproduces all of the hydrogen-like and helium-like emission features observed in the soft X-ray spectrum of NGC 4151 {\it in detail} and confirms the correspondence between the soft X-ray emission in NGC 4151 and NGC 1068. NGC 4151 shows somewhat lower individual ionic column densities than those measured for NGC 1068 indicating that the material in the ionization cones of NGC 4151 may be somewhat more dense than the material in the ionization cones of NGC 1068.

\end{abstract}


\section{Introduction}
\label{1}

Prior to the advent of \xmm ~and \cha, the origin of the soft X-ray emission from heavily obscured AGN remained very uncertain. Recently, however, observations with \xmm ~and \cha ~have revolutionized our understanding of heavily obscured AGN (for a complete description of \xmm, \cha ~and their relative capabilities see Jansen \etal 2001, Weisskopf \etal 2002 and references therein). The combination of the high spatial resolution afforded by \cha, the large collecting area of \xmm ~and the high spectral resolution X-ray spectroscopy capabilities (in the form of X-ray grating and CCD instruments) of both missions across the 0.1-12 keV band is beginning reveal many new details of the X-ray emission from heavily obscured AGN, particularly with regard to the soft X-ray emission from these objects. \cha ~observations of several heavily obscured AGN have revealed considerable extended soft X-ray emission in each source, that is tightly correlated with the O[III]($\lambda$5007) emission observed with the {\it Hubble Space Telescope} (HST; NGC 1068 - Young \etal 2001; Mrk 3 - Sako \etal 2000; NGC 4151 - Ogle \etal 2000; Circinus - Sambruna \etal 2001). In several of these sources ({\it e.g.} NGC 1068 - Young \etal 2001), the extended X-ray emission and the O[III]($\lambda$5007) emission clearly traces the spiral arms and other large-scale structures out to a considerable distance from the active nucleus of the galaxy. Similarly recent spectral observations of these AGN with the X-ray grating instruments on \cha ~(NGC 1068 - Brinkman \etal 2002, Ogle \etal 2002; Mrk 3 - Sako \etal 2000; NGC 4151 - Ogle \etal 2000; Circinus - Sambruna \etal 2001) and \xmm ~(Kinkhabwala \etal 2002) have revealed soft X-ray spectra dominated by prominent emission lines from a range of light elements ({\it e.g.} C, N, O, Ne etc). These observations constitute a growing body of evidence that suggests that the soft X-ray emission from heavily obscured Seyferts is largely composed of soft X-ray emission lines that originate in gas that is photoionized (and photoexcited) by the central AGN.

The Seyfert 1.5 galaxy NGC 4151 was identified as an X-ray source over thirty years ago  (Gursky \etal 1971). One of the brightest AGN accessible in the X-ray band, NGC 4151 has been extensively studied by missions such as \exo, \gin, \ro, \asca, \cgro ~and more recently \xte, \bep, \cha ~and \xmm. This observational focus has revealed that the spectrum of NGC 4151 from 0.1-100 keV  is comprised of a complex mixture of emission and absorption components, probably originating in a variety of locations from the innermost parts of the putative accretion disk in NGC 4151, out to the extended narrow-line region of the galaxy. Despite the relatively poor quality of the spectra, previous observations (particularly observations with \asca) did tentatively reveal several unresolved soft X-ray features, suggesting that the soft X-ray emission was somewhat more complex than had been previously suggested. The most prominent feature found with \asca ~was at an energy of $\sim$0.88 keV and was tentatively associated with O VIII Ly$\alpha$ RRC (Weaver \etal 1994; Griffiths \etal 1998). The detailed nature of the soft X-ray emission in NGC 4151 has only recently been revealed through a \cha ~HETG observation (Ogle \etal 2000, Yang \etal 2001). This observation demonstrated that the soft X-ray spectrum was composed almost entirely of emission lines and RRC, with little or no detected continuum emission of any form (and hence no absorption features). Utilizing the excellent spatial resolution of \cha, Ogle \etal (2000) resolved $\sim$70\% of the soft X-ray emission as extended emission, originating up to 11.5\arcs ~from the AGN, that demonstrated an excellent correspondence with the bi-conical O[III]($\lambda$5007) emission previously imaged with the HST (Evans \etal 1993). The \cha ~observation was even capable of resolving the spatial profiles of several of the brightest emission lines (including the iron K$\alpha$ emission line) showing that the individual line emission was genuinely extended along the direction of the optical ionization cones. The \cha ~observation dramatically confirmed that the hard X-ray spectrum of NGC 4151 is strongly cut-off below ~2 keV and that the soft excess is not predominantly produced by electron scattering of the hard X-ray continuum. The \cha ~analysis finds evidence for both photoionized and collisionally ionized (thermal) gas in the ENLR, however many of the details of the ionized medium remained elusive, due to the large number (33 in total) and low significance of the features detected. 

This paper is organized as follows. Section~\ref{2} details the \xmm ~observations of NGC 4151. Section~\ref{3} presents the \xmm ~RGS spectra and discusses the spectral modelling in terms of emission from an ionization cone. Section~\ref{4} compares the RGS spectra of NGC 4151 with the soft X-ray spectra of NGC 1068 and Mrk 3. Section~\ref{5} presents a preliminary look at the correspondence between the \xmm ~RGS and the \xmm ~EPIC soft X-ray spectra of NGC 4151.

\section{The \xmm ~observations}
\label{2}

NGC 4151 was observed with \xmm ~five times during December 21$^{st}$-23$^{rd}$, 2000. The RGS instruments were active for all five observations and the data were processed with the standard RGS pipeline processing chains incorporated in the \xmm ~SAS v5.4.1. Dispersed source spectra and background spectra were extracted with the automated RGS extraction tasks in the SAS. The SAS filters the RGS events in dispersion-channel vs CCD-pulse-height space to separate the data from the different spectral orders. Investigation of the individual spectra indicate that the 2$^{nd}$ order RGS spectra contain a negligible amount of information (a count rate of $\leq$10\% of that in the 1$^{st}$ order spectra, despite having $\sim$60\% of the effective area below 10\AA) and as a result only the 1$^{st}$ order spectra are presented here. The resulting spectra have a total effective exposure time of $\sim$128 ks and $\sim$125 ks for the RGS 1 and 2 instruments respectivly. A previous investigation of the \xmm ~EPIC spectra confirms that the soft X-ray spectrum of NGC 4151 remained relatively constant for the duration of the observation (Schurch \etal 2003). The spectra from the five observations were co-added to produce a single high signal-to-noise source spectrum and a single background spectrum for each RGS instrument. The full spectra and individual lines were analyzed with both maximum likelihood and $\chi^2$ minimization techniques. For the maximum likelihood analysis the spectra were left unbinned however, for the $\chi^2$ minimization techniques (which, in general, are not particularly well-suited to fitting the sharp features observed in this and similar high resolution soft X-ray spectra) the spectra were binned to a minimum of 20 counts per spectral channel. Investigation of the spatial extent of the X-ray emission observed by \xmm ~reveals that the soft X-ray emission is not significantly extended on scales greater than the mirror PSF, consistent with the details of the previously reported \cha ~observation of NGC 4151 (Ogle \etal 2000).

\section{Modelling the RGS spectra}
\label{3}

\begin{figure*}
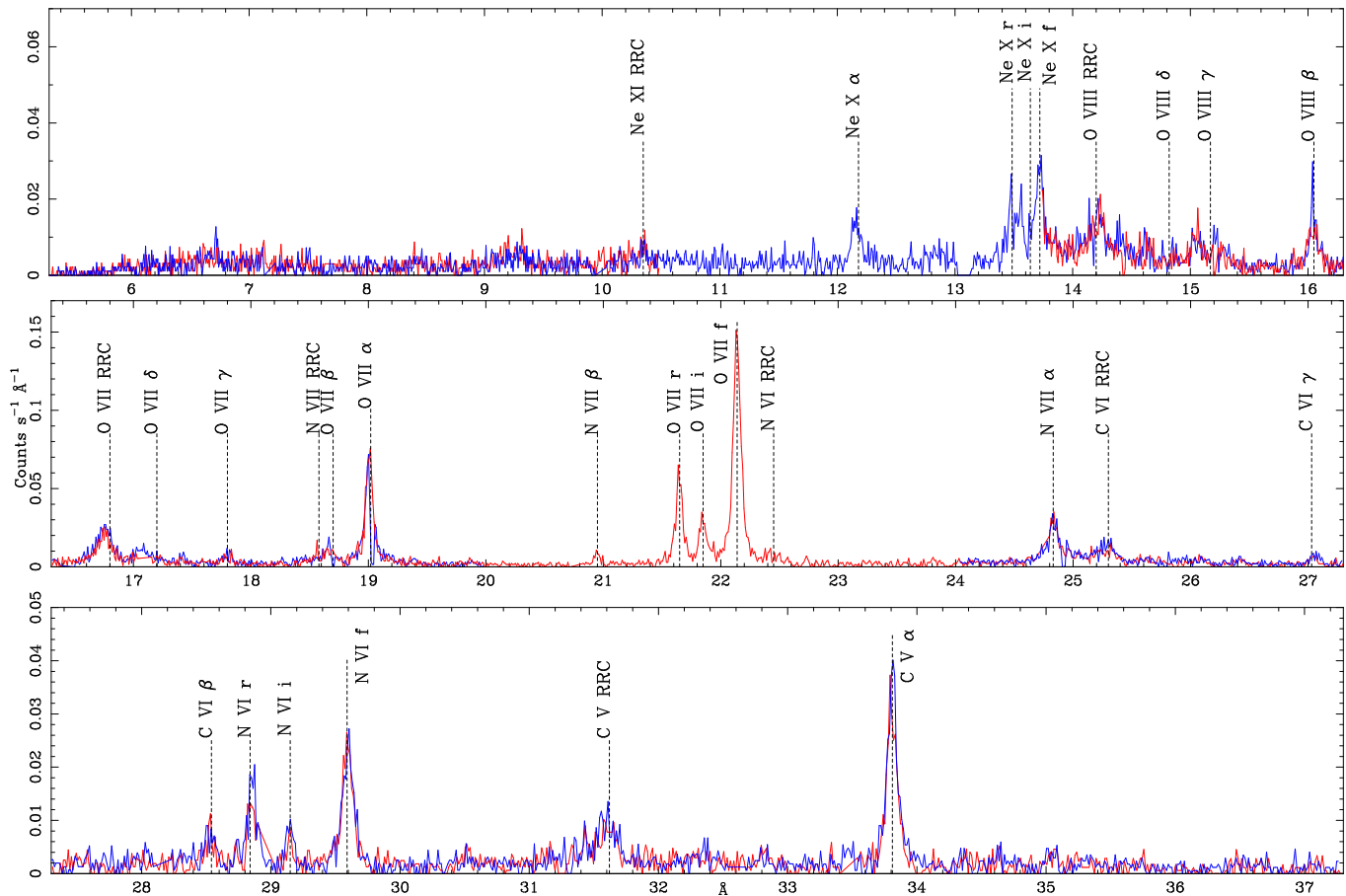

\centering
\begin{minipage}{180 mm}
\centering
\vbox{
\includegraphics[height=18.0 cm, angle=270]{rgs1.ps}
\includegraphics[height=18.0 cm, angle=270]{rgs2.ps}
\includegraphics[height=18.0 cm, angle=270]{rgs3.ps}
}
\caption{RGS 1 (red) and RGS 2 (blue) spectra of NGC 4151. The spectra here are unbinned. Spectral discontinuities are caused by chip gaps in the CCD arrays and the effects of the damaged RGS CCDs. Unambiguously identified hydrogen-like and helium-like emission lines are labeled along with the clearly identified RRC.}
\label{RGS1}
\end{minipage}
\end{figure*}

Figure~\ref{RGS1} shows the \xmm ~RGS spectra from the combined observations of NGC 4151. The spectrum is completely dominated by line emission and radiative recombination continua (RRC). Line emission is clearly detected from hydrogen-like and helium-like ionization states of neon, oxygen, nitrogen and carbon and in many cases the line emission associated with these ions is not limited to the Ly$\alpha$ (2p-1s) transition but includes higher order resonance transitions (np-1s) labeled $\beta$-$\varepsilon$ for the different hydrogen-like and helium-like species respectively. Much of the unidentified line emission below $\sim$16\AA ~can be associated with a multitude of unresolved iron L-shell emission lines, along with some contribution from hydrogen-like and helium-like silicon and magnesium. The spectra show no significant continuum emission and no fluorescent line emission (there may be a marginal detection of silicon K$\alpha$ fluorescence at $\sim$7.1\AA, however without a good knowledge of the iron L-shell emission in this region it is not possible to measure the line cleanly). 

\subsection{The emission lines}
\label{3-1}

The spectrum is remarkably similar to the soft X-ray spectrum of the archetypal bright Seyfert 2 galaxy, NGC 1068, reported by Kinkhabwala \etal (2002). Both sources show strong emission line triplets from helium-like nitrogen and oxygen, with a similar pattern of resonance, intercombination and forbidden line strengths. Following the analysis of the \xmm ~RGS observation of NGC 1068 presented in Kinkhabwala \etal (2002), line fluxes, centroid energies and line widths were determined for the bright, unambiguously identified, emission lines observed in the spectra. The spectral fits quoted here for individual strong line features were performed with $\chi^2$ minimization techniques (and correspondingly the data were binned to a minimum of 20 counts per spectral channel). An examination of the fit properties using maximum likelihood fitting techniques for individual lines reveals that the best-fit parameters are not significantly different using these statistical techniques in comparison with the $\chi^2$ fits, as expected given the bright nature of the features examined here. The intrinsic profiles of the emission lines are represented by a line model with a Gaussian profile (we caution, however, that the original line profile may in fact not be Gaussian, although it does appear to be a good approximation for the lines presented here) which allows the determination of accurate line centroid energies, fluxes and, in some cases, line widths. The results of fitting this profile to the brightest emission lines are given in Table~\ref{RGSGGLfit}. 

\begin{table*}
\centering
\begin{minipage}{180 mm} 
\caption{Line properties of individual emission lines observed in the RGS spectra of NGC 4151}
\centering
\begin{tabular}{lccccc}
 \\
Line ID & $\lambda_{Obs}$ & $\lambda_{Rest}^{a}$ & Line Shift & $\sigma_{Obs}$ & Flux \\
        & {\small \AA~} & \AA~ & {\small km s$^{-1}$} & {\small $\times$10$^{-2}$ \AA} & {\small $\times10^{-4}$ Photons s$^{-1}$ cm$^{-2}$} \\ \hline
\vspace{-0.35cm}\\
\vspace{0.1cm} Ne {\small X} $\alpha$ & 12.13$^{+0.01}_{-0.01}$ & 12.134 & -120$^{+250}_{-250}$ & 2.6$^{+1.0}_{-0.9}$ & 0.42$^{+0.05}_{-0.05}$ \\
\vspace{0.1cm} Ne {\small IX} r & 13.43$^{+0.02}_{-0.01}$ & 13.447 & -430$^{+490}_{-260}$ &  2.4$^{+1.9}_{-1.1}$ & 0.47$^{+0.05}_{-0.05}$ \\
\vspace{0.1cm} Ne {\small IX} i & 13.53$^{+0.01}_{-0.02}$ & 13.550 & -470$^{+240}_{-340}$ &  2.0$^{+1.6}_{-1.2}$ & 0.46$^{+0.15}_{-0.07}$ \\
\vspace{0.1cm} Ne {\small IX} f & 13.68$^{+0.01}_{-0.02}$ & 13.697 & -310$^{+220}_{-180}$ &  & 1.03$^{+0.1}_{-0.09}$  \\ \hline
\vspace{-0.35cm}\\
\vspace{0.1cm} O {\small VIII} $\beta$ & 16.00$^{+0.02}_{-0.02}$ & 16.006 & -150$^{+230}_{-230}$ & 3.1$^{+0.6}_{-0.6}$ & 0.37$^{+0.04}_{-0.04}$ \\
\vspace{0.1cm} O {\small VIII} $\alpha$ & 18.96$^{+0.01}_{-0.01}$ & 18.969 & -210$^{+130}_{-130}$ &  & 1.60$^{+0.10}_{-0.10}$ \\
\vspace{0.1cm} O {\small VII} $r$ & 21.59$^{+0.01}_{-0.01}$ & 21.603 & -230$^{+140}_{-140}$ &  & 1.19$^{+0.09}_{-0.09}$ \\
\vspace{0.1cm} O {\small VII} $i$ & 21.79$^{+0.01}_{-0.01}$ & 21.798 & -230$^{+120}_{-130}$ &  & 0.75$^{+0.07}_{-0.07}$ \\
\vspace{0.1cm} O {\small VII} $f$ & 22.08$^{+0.01}_{-0.01}$ & 22.101 & -300$^{+110}_{-110}$ & 0.68$^{+0.09}_{-0.09}$ & 3.89$^{+0.17}_{-0.19}$ \\ \hline
\vspace{-0.35cm}\\
\vspace{0.1cm} N {\small VII} $\beta$ & 20.89$^{+0.01}_{-0.01}$ & 20.910 & -230$^{+160}_{-150}$ & 0.7$^{+0.5}_{-0.4}$ & 0.22$^{+0.04}_{-0.04}$ \\
\vspace{0.1cm} N {\small VII} $\alpha$ & 24.77$^{+0.01}_{-0.01}$ & 24.781 & -190$^{+110}_{-110}$ &  & 0.81$^{+0.09}_{-0.08}$ \\
\vspace{0.1cm} N {\small VI} $r$ & 28.75$^{+0.01}_{-0.01}$ & 28.792 & -350$^{+110}_{-110}$ &  & 0.45$^{+0.07}_{-0.06}$ \\
\vspace{0.1cm} N {\small VI} $i$ & 29.09$^{+0.02}_{-0.02}$ & 29.075 & 30$^{+250}_{-160}$ &  & 0.36$^{+0.06}_{-0.06}$ \\
\vspace{0.1cm} N {\small VI} $f$ & 29.51$^{+0.01}_{-0.01}$ & 29.531 & -290$^{+100}_{-110}$ &  & 1.15$^{+0.12}_{-0.12}$ \\ \hline
\vspace{-0.35cm}\\
\vspace{0.1cm} C {\small VI} $\beta$ & 28.43$^{+0.01}_{-0.01}$ & 28.466 & -390$^{+130}_{-130}$ & 0.9$^{+0.3}_{-0.3}$ & 0.37$^{+0.05}_{-0.05}$ \\
\vspace{0.1cm} C {\small VI} $\alpha$ & 33.69$^{+0.01}_{-0.01}$ & 33.736 & -440$^{+90}_{-90}$ &  & 1.79$^{+0.16}_{-0.16}$ \\ \hline
\vspace{-0.35cm} \\
\multicolumn{3}{l}{$^{a}$ Vainshtein \& Safronova, 1978} \\
\end{tabular}
\label{RGSGGLfit}
\end{minipage}
\end{table*}

The measured fluxes take account of the Galactic column density (2.1$\times10^{20}~cm^{-2}$; Warwick \etal 1995) using the {\bf tbabs} model in XSPEC, and the line shifts are measured with respect to the source redshift of z=0.00332. The systematic uncertainty on measuring the centroid line energies of the lines is estimated to be $\sim$8$\times10^{-3}$\AA ~(which corresponds to about 150 km s$^{-1}$; {\it XMM} Users Handbook v2.1, 2002). The quoted errors are 90\% confidence levels as defined by a $\Delta$$\chi$$^{2}$=2.71 criterion ({\it i.e.} assuming one interesting parameter) and include the systematic uncertainty added in quadrature. Most of the measured line centroid energies are consistent with a mild blueshift, of between $\sim$100-1000 km s$^{-1}$, with respect to the systemic velocity of the lines. This implies a line-of-sight outflow velocity for the ionized material of $\ls$1000 km s$^{-1}$, consistent with the line shifts of the absorption lines observed in the UV spectrum of NGC 4151 ($\sim$600 km s$^{-1}$; Kriss \etal 1995). Many of the observed line widths are largely unconstrained, with the spectral fitting only yielding upper limits (these are not quoted in Table~\ref{RGSGGLfit}); however the seven line widths that are constrained by the spectral fitting appear to show a bimodal distribution with the shorter wavelength lines having an average line width of $\sigma$$\sim$0.025 \AA ($\sim$1000 km s$^{-1}$ FWHM) and the longer wavelength lines having an average line width of $\sim$0.005 \AA ($\sim$200 km s$^{-1}$ FWHM). The observed line widths imply that the intrinsic velocity dispersion of the outflowing material is $\ls$1000 km s$^{-1}$ which is somewhat lower than the line widths of the absorption lines observed in the UV (1000-2500 km s$^{-1}$; Kriss \etal 1995). The bimodal distribution of the line widths as a function of the wavelength of the line is suggestive of a scenario in which individual clouds of emitting material close to the source of the ionizing continuum are moving more rapidly than clouds further from the continuum source. The large majority of the soft X-ray flux of NGC 4151 as measured by \cha ~is not extended on scales larger than $\sim$2\arcs, demonstrating that the reported extended soft X-ray flux will not contribute considerably to the line widths measured in the RGS spectrum.

\subsection{Temperature diagnostics with Radiative Recombination Continua}
\label{3-2}

The radiative recombination continua (RRC) identified in the RGS spectra are the result of a free electron recombining directly into a bound state of an ion. Recombination to the ground state represents the dominant transition and is the only transition considered during the spectral modelling. The free electrons can have any energy above the recombination threshold and hence the feature produced is neither symmetrical or narrow. The energy of the electrons prior to the recombination process is dependent upon the electron temperature of the emitting plasma and hence the width of the RRC features provides a direct measurement of the recombining electron temperature, $T_{\mathrm e}$ (Liedahl \& Paerels, 1996; Liedahl 1999). RRC originating in hot plasmas are broad and often subtle features however, RRC from cool plasmas are narrow, prominent, features. The velocity dispersion of the emitting material will affect the RRC in exactly the same manner as it affects the emission lines discussed previously, however this effect is negligible here due to the low velocities of the ions involved compared to that of the electrons. RRC from hydrogen-like and helium-like ionization states of Ne, O, N and C are clearly identifiable in Figure~\ref{RGS1}. Electron temperatures were determined for the bright, unambiguously identified, RRC (Table~\ref{RRCtemps}) using the RRC profiles generated by the 'ionization cone' model discussed in Section 3.4 fit, using $\chi^2$ minimization techniques, to the individual RRC. Again, an examination of the fit properties using maximum likelihood fitting techniques for individual features reveals that the best-fit parameters are not significantly different using these statistical techniques in comparison with the $\chi^2$ fits. The derived electron temperatures imply electron temperatures of $\sim$2-4$\times10^{4}$ K. The association of high-ionization emission lines ({\it e.g.} O VIII, Ne X etc; identified in the previous section) with such a low temperature plasma is strongly indicative of a plasma whose ionization state is governed by photoionization as opposed to collisional ionization.

\begin{table}
\centering
\begin{minipage}{80 mm} 
\caption{Electron temperatures (in eV) from RRC}
\centering
\begin{tabular}{lc}
 \\
RRC & $T_{\mathrm e}$ \\ \hline
\vspace{-0.35cm} \\
\vspace{0.1cm} C V & 2.5$^{+1.5}_{-0.5}$ \\
\vspace{0.1cm} C VI & 4.0$^{+1.0}_{-1.0}$ \\
\vspace{0.1cm} N VI & 3.0$^{+3.0}_{-1.0}$ \\
\vspace{0.1cm} N VII & 4.0$^{+2.0}_{-1.5}$ \\
\vspace{0.1cm} O VII & 4.0$^{+0.5}_{-0.5}$ \\
\vspace{0.1cm} O VIII & 6.0$^{+1.0}_{-2.0}$ \\ \hline
\vspace{-0.35cm} \\
\end{tabular}
\label{RRCtemps}
\end{minipage}
\end{table}

\subsection{Plasma diagnostics with helium-like triplets}
\label{3-3}

The three most prominent X-ray emission lines from helium-like ions correspond to transitions from the n=2 shell to the n=1 shell and are the resonance (1s2p $^{1}$P$_{1}$$\rightarrow$1s$^{2}$ $^{1}$S$_{0}$), intercombination (1s2p $^{3}$P$_{2,1}$$\rightarrow$1s$^{2}$ $^{1}$S$_{0}$) and forbidden (1s2s $^{3}$S$_{1}$$\rightarrow$1s$^{2}$ $^{1}$S$_{0}$) line transitions (Pradhan \etal 1981). Ratios of the measured line fluxes for the recombination, intercombination and forbidden lines from helium-like ions provide a diagnostic of the electron density and temperature in hot plasmas. Gabriel \& Jordan (1969) were the first to recognise the importance of the helium-like triplet lines and used line strength ratios to analyze the solar corona spectra in detail. The ratio of the intensity of the forbidden line to the intensity of the intercombination line (R, where R=f/i) is sensitive to the electron density of the plasma and the ratio of the sum of the intensities of the forbidden and intercombination lines to the intensity of the resonance line [G, where G=(i+f)/r] is sensitive to the electron temperature of the plasma. For a detailed discussion of the manner in which the ratios depend on the electron density and the plasma temperature see Pradhan \etal (1981), Porquet \& Dubau (2000), Porquet \etal (2002) and references therein.

\begin{table}
\centering
\begin{minipage}{80 mm} 
\caption{Helium-like triplet plasma diagnostic ratios}
\centering
\begin{tabular}{lcc}
 \\
Triplet ID & R & G \\ \hline
\vspace{-0.35cm} \\
\vspace{0.1cm} Ne{\small IX} & 2.3$^{+0.7}_{-0.7}$ & 3.2$^{+1.2}_{-0.9}$ \\
\vspace{0.1cm} O{\small VII} & 3.4$^{+0.4}_{-0.3}$ & 3.9$^{+0.5}_{-0.5}$ \\
\vspace{0.1cm} N{\small VI} & 3.0$^{+1.1}_{-0.8}$ & 3.4$^{+1.0}_{-0.8}$ \\ \hline
\vspace{-0.35cm} \\
\end{tabular}
\label{HElinerats}
\end{minipage}
\end{table}

The R and G ratios for the Ne, O and N triplets observed in the RGS spectrum of NGC 4151 are given in Table~\ref{HElinerats}. The G-values imply an electron temperature of $\ls$$10^{5}$ K, consistent with the electron temperatures implied from the properties of the observed RRC and consistent with photoionization being the dominant ionization mechanism in the plasma (Porquet \& Dubau, 2000). The R-values imply electron densities of $\sim$10$^{8}$-10$^{9}$ cm$^{-3}$ in nitrogen and oxygen and $\sim$10$^{10}$-10$^{11}$ cm$^{-3}$ from neon (Porquet \& Dubau, 2000). In particular the ratios calculated from the oxygen triplet (which is both the strongest and the least confused of the helium-like triplets) strongly imply a photoionized plasma however, extracting detailed values of density and temperature for the emitting material is complicated by the possibility that a strong radiation field ({\it i.e.} photoexcitation) can mimic the effect that a relatively high density has on the line strengths. In such a situation the electron density, and to a lesser extent, the electron temperature can be overestimated (Porquet \etal 2002). We note that the expected contribution of blended dielectronic satellite lines to these ratios has not been taken into account in the values quoted here. For the N and O triplets this contribution is negligible; however the Ne triplet is significantly affected. For a high temperature, collisionally dominated, plasma the additional contribution can result in an error of $\sim$1\% and $\sim$9\% for the R and G ratios respectively, however for a photoionized plasma, where the temperatures are several orders of magnitude lower, this effect can be considerably greater (Porquet \etal 2002).

\begin{figure}
\centering
\begin{minipage}{80 mm} 
\centering
\includegraphics[width=7 cm, angle=0]{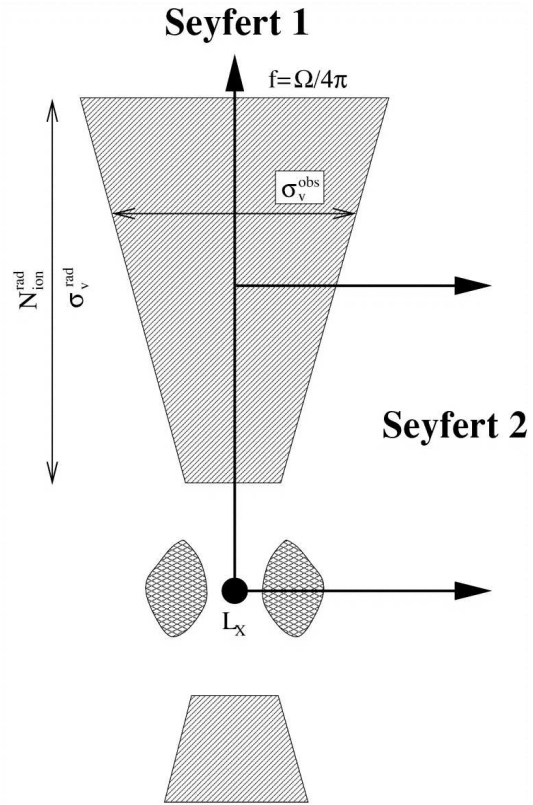}
\caption{Schematic cartoon of the `ionization cone' model (not to scale). The central nuclear components (black hole and accretion disk), denoted by the black spot, are surrounded by a dense absorbing torus of neutral dust and gas. The optically thin material perpendicular to the plane of the absorbing torus sees the direct continuum from the X-ray source and becomes strongly ionized forming ionization cones. The line emission from these cones is observable from Seyfert 2 galaxies due to the suppression of the direct continuum, whereas the ionization cones in Seyfert 1 galaxies are only seen through their absorption of the direct continuum (Kinkhabwala \etal 2002).} 
\label{KIN02_fig6}
\end{minipage}
\end{figure}

\subsection{The ionization cone model}
\label{3-4}

Given the similarity between the RGS spectra of NGC 1068 and NGC 4151, we utilize the `ionization cone' model developed by Kinkhabwala \etal (2002) in their analysis of the RGS spectrum of NGC 1068, to fit the RGS spectrum of NGC 4151. This model, which was incorporated into XSPEC for ease of spectral fitting, reproduces the hydrogen-like and helium-like line emission, and the associated RRC, from an irradiated cone of optically thin plasma ({\it c.f.} the ionization cones observed in O[III] by HST) shown schematically in Figure~\ref{KIN02_fig6}. The model includes the effects of both photoionization and, crucially, photoexcitation. Photoexcitation is the process by which the absorption of a photon with insufficient energy to ionize the atom, results in the transition of an inner-shell electron to a higher energy level within the atom. The excited electron will rapidly undergo one or more radiative transitions, decaying back to its original energy state. For each atomic species, photoexcitation is expected to enhance the intensity of the higher order resonance transition lines relative to the 1$^{st}$ order resonance line. Photoexcitation was observed to play a significant role in the spectrum of NGC 1068 where the Ly$\beta$, Ly$\gamma$ and Ly$\delta$ lines were significantly stronger than predicted by models based on pure photoionization or collisional ionization. 

Here, as with NGC 1068, the plasma is irradiated by the unabsorbed intrinsic power-law continuum, the slope of which in the case of NGC 4151 is fixed at $\Gamma$=1.65, based on the analysis of the hard X-ray \xmm ~EPIC spectrum presented in Schurch \etal (2003). The luminosity of the central source, L$_{X}$, and the covering fraction of the cone, f, are fixed at L$_{X}$=6$\times10^{43}$ erg s$^{-1}$ and f=0.1, respectively. The modelling includes both Galactic column density (2.1$\times10^{20}$ cm$^{-2}$; Warwick \etal 1995) using the {\bf tbabs} model in XSPEC and a source redshift of z=0.00332. The electron temperatures for each atomic species were fixed at the temperatures found from the analysis of the RRC (Table~\ref{RRCtemps}), where possible, and the remaining RRC temperatures were fixed at the temperatures identified for NGC 1068 by Kinkhabwala \etal (based on the similarity between the observed bright RRC in NGC 4151 and the corresponding RRC in NGC 1068). The remaining free parameters of the model are the transverse velocity distribution of the lines, $\sigma^{obs}_{v}$, the radial velocity distribution of the lines, $\sigma^{rad}_{v}$, and the radial column density of each individual atomic species, N$^{rad}_{ion}$. The modelling presented in the next section includes a significant contribution from iron L-shell emission lines. Iron L-shell emission, known to be present in the soft X-ray spectrum of both Mrk 3 and NGC 1068, will contribute significantly between $\sim$9-16\AA (0.75-1.4 keV) and will results in an artificially high measure of the ionic column densities of Neon and Magnesium if it is not incorporated in the modelling.

\subsection{Fitting the ionization cone model to the RGS spectra}
\label{3-5}

\begin{figure*}
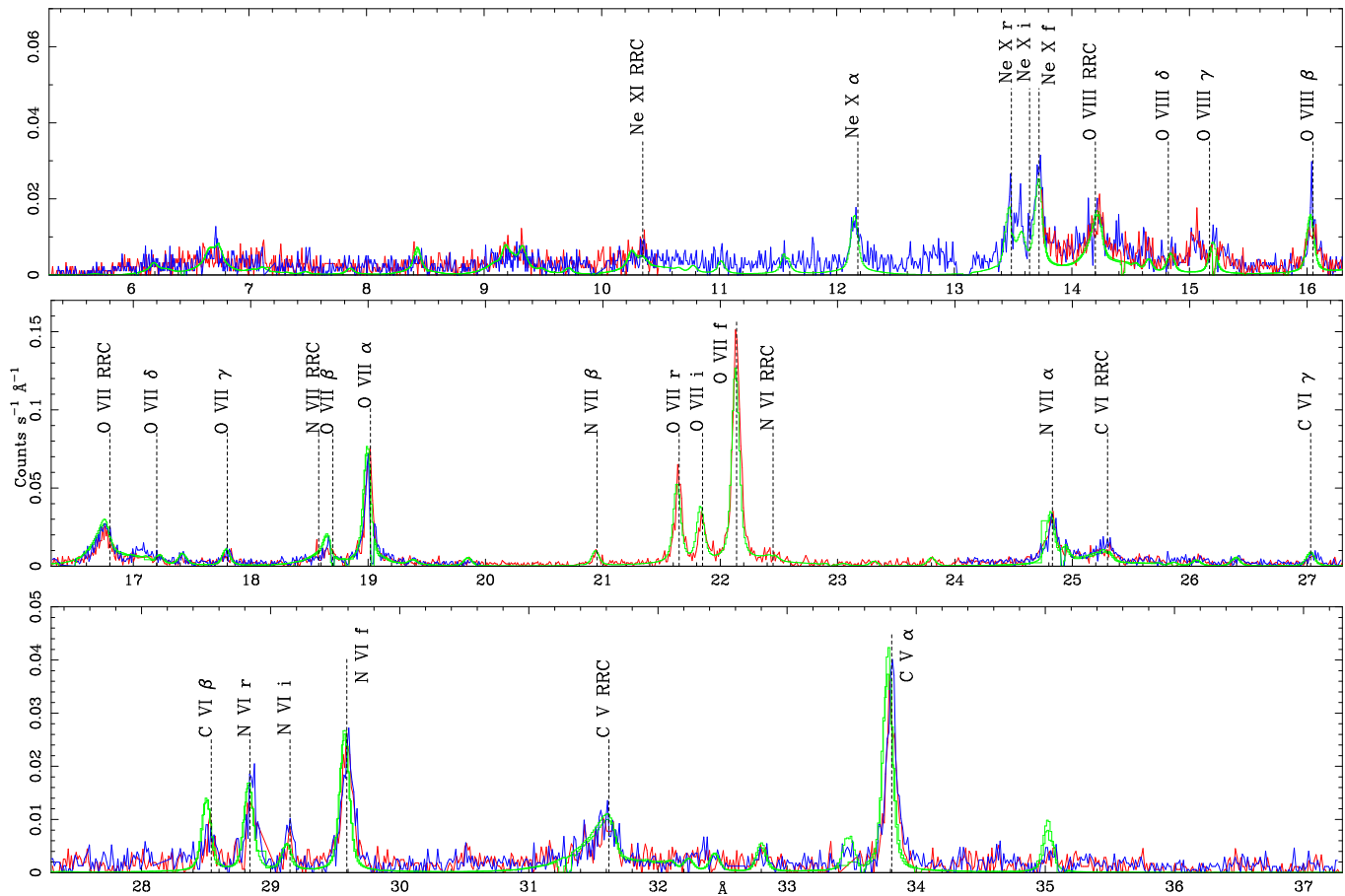

\centering
\begin{minipage}{180 mm}
\centering
\vbox{
\includegraphics[height=18.0 cm, angle=270]{rgs1+mod.ps}
\includegraphics[height=18.0 cm, angle=270]{rgs2+mod.ps}
\includegraphics[height=18.0 cm, angle=270]{rgs3+mod.ps}
}
\caption{Best-fit of the ionization cone model to the RGS spectra of NGC 4151. The unbinned RGS 1 and RGS 2 spectra are shown in red and blue along with the best-fit model in green. The best-fit model parameters are given in Table~\ref{ioncfit}. Unambiguously identified hydrogen-like and helium-like emission lines and RRC are labeled in a similar fashion to that in Figure~\ref{RGS1}.}
\label{RGS1mod}
\end{minipage}
\end{figure*}

\begin{table}
\centering
\begin{minipage}{80 mm} 
\caption{Best-fit parameters for the `ionization cone' model fit to the RGS spectra of NGC 4151}
\centering
\begin{tabular}{lcc}
 \\
ion & N$^{rad}_{ion}$ ({\small cm$^{-2}$}) & k$T_{\mathrm e}$ ({\small eV}) \\ \hline
\vspace{-0.35cm} \\
\vspace{0.1cm} C V   & 2.9$\times10^{17}$ & 2.5 \\
\vspace{0.1cm} C VI  & 4.0$\times10^{17}$ & 4.0 \\
\vspace{0.1cm} N VI  & 1.4$\times10^{17}$ & 3.0 \\
\vspace{0.1cm} N VII & 2.4$\times10^{17}$ & 4.0 \\
\vspace{0.1cm} O VII & 7.6$\times10^{17}$ & 4.0 \\
\vspace{0.1cm} O VIII& 7.9$\times10^{17}$ & 6.0 \\
\vspace{0.1cm} Ne IX & 4.6$\times10^{17}$ & 4.0$^{a}$ \\
\vspace{0.1cm} Ne X  & 4.7$\times10^{17}$ & 4.0$^{a}$  \\
\vspace{0.1cm} Mg XI & 2.2$\times10^{17}$ & 4.0$^{a}$  \\
\vspace{0.1cm} Mg XII& 4.0$\times10^{17}$ & 4.0$^{a}$  \\
\vspace{0.1cm} Si XIII & 6.2$\times10^{17}$ & 4.0$^{a}$  \\
\vspace{0.1cm} Si XIV& 1.1$\times10^{18}$ & 4.0$^{a}$  \\
\vspace{-0.35cm} \\ \hline
\multicolumn{3}{l}{{\small $^{a}$ Unconstrained parameter. Value}} \\
\multicolumn{3}{l}{{\small fixed at the value used in}} \\
\multicolumn{3}{l}{Kinkhabwala \etal (2002).} \\
\end{tabular}
\label{ioncfit}
\end{minipage}
\end{table}

The ionization cone model was fitted to the RGS spectra in XSPEC for simplicity. We note that the $\chi^2$ minimization techniques, used earlier in the paper to fit specific bright spectral features, are, in general, not well-suited to fitting wide-band X-ray grating spectra which often incorporate many additional low-significance spectral features. With this in mind, the ionization cone model is fit to the RGS spectra of NGC 4151 using maximum likelihood fitting techniques. The complexity of this specific model makes the accurate determination of individual parameter errors problematic. Despite this caveat, the model provides a ``good fit'' to all the hydrogen-like and helium-like line emission and RRC observed in the RGS spectra. Here, the term ``good fit'' here refers only to a set of parameters that correctly reproduces the distinct features observed in the spectrum ({\it i.e.} it is not a statistical measure of the goodness of fit of the model). Because of this, statistical errors are not quoted for either the column densities or the velocity measurements. Figure~\ref{RGS1mod} shows the best-fit of the ionization cone model to all the hydrogen-like and helium-like emission lines and RRC in the RGS spectrum of NGC 4151. The best-fit radial ionic column densities and the associated RRC electron temperatures are given in Table~\ref{ioncfit}; $\sigma^{obs}_{v}$ has a best-fit value of $\sim$400 km s$^{-1}$ and $\sigma^{rad}_{v}$ has a best-fit value of $\sim$330 km s$^{-1}$. We do not quote any specific best-fit values for the ionic column densities of iron responsible for the iron L-shell emission (Fe {\small XVI-XXIII}) due to the unresolved nature of the specific iron L-shell lines. Typically the ionic iron column densities are of the order of $\sim$few $\times$10$^{16}$ cm$^{-2}$.

Both the velocity measurements are in reasonable agreement with the average line widths and line shifts measured from the individual line centroid energies. Kinkhabwala \etal (2002) quote, based on the NGC 1068 data, an estimate for the uncertainties on the individual model parameters of approximately a few tens of percents for the ionic column densities and approximately a factor of 2 for the transverse and radial velocities. Here, the data are of somewhat worse statistical quality than the NGC 1068 spectra and so we expect the parameters to be less well constrained. NGC 1068 is a factor of $\sim$3 times brighter than NGC 4151 and hence the errors on the fit parameters from the NGC 4151 spectrum will be a factor of $\sim$2 times greater than the errors estimated for the NGC 1068 fit parameters. Using a fractional ionic abundance of 0.5 for O VIII, the column density of O VII quoted in Table~\ref{ioncfit} yields an equivalent hydrogen column density of N$_{H}$$\gs$2$\times10^{21}$ cm$^{-2}$.

\section{A comparison with NGC 1068 and Mrk 3}
\label{4}

The archetypal Seyfert 2 galaxies, NGC 1068 and Mrk 3 have both been the targets of recent X-ray grating observations. The soft X-ray emission in both of these sources is extremely similar to the soft X-ray emission from NGC 4151; it is extended on scales of hundreds of parsecs, shows a strong correlation with the optical ionization cones identified in HST O[III]($\lambda$5007) images (Young \etal 2001; Sako \etal 2000) and the soft X-ray spectra are all dominated by similar emission lines and RRC.  

\subsection{The comparison with the \xmm ~RGS spectrum of NGC 1068}
\label{4-1}

The soft X-ray \xmm ~RGS spectrum of NGC 4151 is remarkably similar to that of NGC 1068. Both sources show strong hydrogen-like and helium-like line emission and RRC from carbon, nitrogen, oxygen, neon, magnesium and silicon. The resemblance between the spectra is apparent in the details of the spectra as well as on a more general level. For example, both sources show similar patterns of resonance, intercombination and forbidden line strengths for the helium-like triplets of nitrogen and oxygen and both show significant (but unresolved) iron L-shell emission. The properties of the detected emission lines are remarkably alike in the two sources, both in terms of the lines present and the scale of their energy shifts. In both sources the prominent emission lines are blue-shifted by $<$1000 km s$^{-1}$ with average line shifts of 305 km s$^{-1}$ and 375 km s$^{-1}$ for NGC 4151 and NGC 1068, respectively (Kinkhabwala \etal 2002). The line widths measured for NGC 1068 do not show the bimodal distribution observed in NGC 4151. However, since only 7 lines have constrained line widths in the NGC 4151 spectrum, compared to 23 lines with constrained widths in the NGC 1068 spectrum, caution should be strongly emphasized in interpreting the validity of the distribution of line widths in NGC 4151. Interestingly the average line width found in the NGC 1068 data ($\sim$1100 km s$^{-1}$ FWHM) is comparable to the average line width of the group of broader lines in the NGC 4151 spectrum ($\sim$1000 km s$^{-1}$ FWHM). It is tempting, based upon a comparison of the helium-like triplet plasma diagnostic ratios (R and G; Section~\ref{3-3}) to infer that the ionization cones in NGC 4151 have a somewhat lower electron temperature and a somewhat higher electron density than those in NGC 1068 (Kinkhabwala \etal 2002). Unfortunatly the situation may not be a clear as cut as this simple comparison suggests, primarily due to the unknown effect of photoexcitation on the diagnostic ratios (discussed previously in Seciton~\ref{3-3}). The best-fit ionic column densities for NGC 1068 were found to be similar to those observed {\it in absorption} for Seyfert 1 galaxies, implying either radially stratified ionization zones or, more likely, a broad distribution of densities at any given radius. An inspection of the radial column densities fit to the RGS spectrum reveals that NGC 4151 appears to have less ionized material in its ionization cones, by a factor of $\sim$3.5 on average. The best-fit ionic electron temperatures, based on the widths of the RRC features, are also extremely similar between NGC 4151 and NGC 1068 (although the comparatively large errors on these temperatures prevents any definite conclusions being drawn regarding this similarity). The ionic column densities are dependent on a combination of factors including, cone length, density, ionization parameter and the filling factor of the material. Lower ionic column densities could be the result of lower ionizing flux or less material in the ionization cones, however a somewhat more plausible explanation, in line with the smaller R-values from the helium-like line triplets, could be that the material in the ionization cones of NGC 4151 is rather more dense than the material in the ionization cones of NGC 1068. Since the recombination rate is dependent on the density of the material, this can result in a smaller ionized fraction for a similar ionizing continuum.

\subsection{The comparison with the \cha ~HETG observation of Mrk 3}
\label{4-2}

The \cha ~HETG observation of Mrk 3 was the first published high resolution soft X-ray spectrum of a Seyfert 2 galaxy and showed unambiguously that the soft X-ray emission in Mrk 3 is composed almost entirely of emission lines and RRC, with little or no continuum emission. The soft X-ray spectrum of Mrk 3 is, again, remarkably analogous to the soft X-ray spectrum of NGC 4151, though with somewhat less statistical significance due to the rapidly decreasing effective area of the HETG instrument at wavelengths above $\sim$9\AA. In both sources there is strong evidence to suggest that the observed emission lines are produced in a photoionized and photoexcited plasma (Sako \etal 2000). The importance of photoexcitation in the spectrum of Mrk 3 is highlighted by the observed weak iron L-shell emission, rather than by the enhanced higher order resonance transitions, rejecting explanations from a recombination dominated or a collisionally dominated plasma. Despite this impressive spectrum, the \cha ~observation of Mrk 3 only partially resolves two of the helium-like line triplets detected in the RGS spectrum of NGC 4151 (oxygen and neon), due to the limited number of photons in each feature, and as a result many of the properties of the plasma are not well constrained. In particular, the resonance and intercombination lines of these triplets cannot be separated and as a result Sako \etal (2000) quote a different diagnostic line ratio ({\it i.e.} (r+i)/f) to the line ratios used here. 

At wavelengths shorter than $\sim$9\AA ~the \cha ~spectrum shows evidence for a reflection dominated continuum along with clear detections of emission lines that are either extremely difficult or impossible to detect with the RGS (given its limited spectral bandpass), namely Fe XXVI, Fe XXV, Fe K$\alpha$, S XVI, S XV, S K$\alpha$, Si XIV, Si XIII, Si K$\alpha$, Mg XII and Mg XI (Sako \etal 2000). This plethora of lines highlights the complimentary capabilities of \xmm ~and \cha ~and strongly supports future observations of bright, heavily obscured AGN with both \cha ~(using the HETG) and \xmm, to obtain the highest possible resolution X-ray spectra over the widest possible energy range. The line ratios calculated from the \cha ~detections of the helium-like silicon and magnesium line triplets ({\it i.e.} Si XIII and Mg XI) are all inconsistent with the expected values for a collisionally ionized plasma and are marginally consistent with the predictions from a photoionized plasma, similar to the conclusions from the triplet ratios for both NGC 4151 and NGC 1068.

\section{A brief comparison with the \xmm ~EPIC spectra}
\label{5}

One of the strengths of \xmm ~is that all the science instruments operate simultaneously during each observation, allowing a direct comparison between the high resolution spectra from the RGS instruments and the lower resolution spectra from the EPIC CCD detectors. This comparison is particularly useful in exposing features in each source spectrum that the other detectors struggle to detect. For example, broad features present in the 0.3-2.5 keV range are not easily identified in RGS spectra, but will be more clearly observed by the EPIC instruments. Conversely, features that appear to be broad in the EPIC spectrum may be resolved into a set of individual narrow lines by the RGS instruments. Conclusions from the process of comparing the EPIC and RGS spectra are complicated by the need for good calibration {\it and cross-calibration} for each instrument involved. 

\begin{figure}
\centering
\begin{minipage}{80 mm} 
\centering
\includegraphics[height=8.0 cm, angle=270]{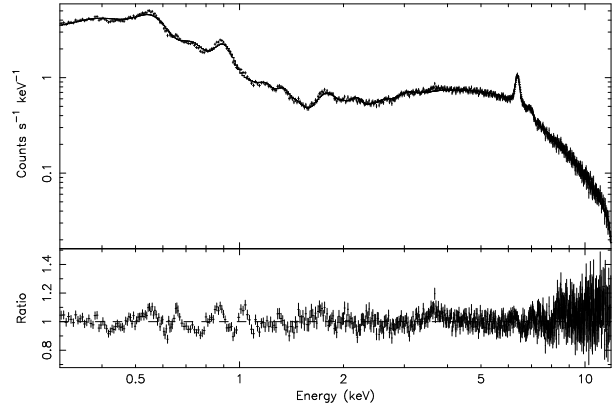}
\caption{{\it Upper Panel}: The 0.3-12 keV EPIC PN spectra of NGC 4151 modelled by a combination of the hard X-ray model developed in Schurch \etal (2003) and the `ionization cone' soft X-ray model presented in Section~\ref{3-4}. {\it Lower Panel}: The ratio of the data to the model prediction.}
\label{EPICRGScomp}
\end{minipage}
\end{figure}

Figure~\ref{EPICRGScomp} shows the full window mode EPIC PN spectra fit with a model that combines the best-fit of the hard X-ray model developed in Schurch \etal (2003) and the best-fit of the `ionization cone' soft X-ray model presented in Section~\ref{3-4}. In addition to the hard X-ray model and the ionization cone model, a silicon K$\alpha$ line was included in the fitting. The energy of the line was fixed at 1.73 keV and the line width was fixed to be the same width as the iron K$\alpha$ line width. The line flux was fixed to be 5\% of the flux contained in the iron K$\alpha$ line, in keeping with the ratio of the iron K$\alpha$ to silicon K$\alpha$ line fluxes measured in the high resolution \cha ~spectrum of Mrk 3. Initially the column densities of the hydrogen and helium-like atomic species were fixed to be the values measured by the RGS data. Key parameters in the hard X-ray model ({\it e.g.} ionization parameter, iron K$\alpha$ line flux etc; see Schurch \etal 2003 for details) were free parameters in the fit, but no contributions from iron L-shell transitions were included in the fit. This highly restricted model does not provide a good fit to the EPIC spectrum, with considerable residuals remaining below 2 keV. Relaxing the constraints of the model by incorporating a contribution from iron L-shell emission, improves the fit considerably and results in similar values for the column densities of iron ionization states from iron XVI-XXIII to those found with the RGS. Despite the improvement, the overall fit remains poor with considerable residuals still remaining across the soft X-ray region. Relaxing the constraints of the model more considerably by allowing the column densities of the hydrogen and helium-like atomic species to be free parameters in the model fit, provides a reasonable fit to the broadband X-ray spectrum of NGC 4151 ($\chi^{2}$=2709 for 2194 d.o.f) however residuals remain in the fit below 2 keV at a level of $\sim$10\% (Figure~\ref{EPICRGScomp}). The best-fit column densities for the hydrogen and helium-like atomic species are typically within a factor of $\sim$3 of the best-fit column densities determined from the RGS modelling. We do note quote these values here because the relatively poor spectral resolution of the EPIC instruments result in considerable degeneracy between the column densities measured for specific atomic species making any specific values at best unreliable and at worst misleading. Despite this caveat, we note that the best-fit column densities of Nitrogen VII, Neon X and Silicon XIV measured in the EPIC data are considerably lower in comparison with those measured in the RGS Data (factors of 3.5, 40, and 16 respectively). The modelling is not significantly improved by the addition of either a soft X-ray power-law (representing a scattered fraction of the underlying continuum) or blackbody emission (to represent the canonical 'soft X-ray excess' often observed in Seyfert 1 galaxies). Considerably more detailed modelling of the RGS spectrum utilizing the additional data from the most recent \xmm ~observations of NGC 4151, coupled with improvements in the relative instrument calibrations, is required before any firm conclusions on the nature of the remaining fit residuals can be determined.

\section{Synopsis}
\label{6}

We present a detailed analysis of the RGS spectrum of NGC 4151 from a total of $\sim$125 ks of \xmm ~observations. The EPIC spectra from these observations are presented in Schurch \etal (2003). The analysis uses the high spectral resolution of the RGS instruments to characterize the nature and origin of the soft X-ray spectrum of NGC 4151.

\begin{itemize}
\item  The soft X-ray spectrum of NGC 4151 (Figure~\ref{RGS1}) is completely dominated by line emission and radiative recombination continua. Line emission, including higher order ($\beta$-$\varepsilon$) resonance transitions, is clearly detected from hydrogen-like and helium-like ionization states of neon, oxygen, nitrogen and carbon. Much of the unidentified line emission below $\sim$16\AA ~can be associated with a combination of iron L-shell emission and lines from hydrogen-like and helium-like silicon and magnesium.

\item The RGS spectra show no significant continuum emission and no fluorescent line emission associated with neutral material.

\item The measured line centroid energies are consistent with a blueshift of between $\sim$100-1000 km s$^{-1}$, with respect to the systemic velocity of NGC 4151. The seven observed line widths that are constrained by the data tentatively suggest a bimodal distribution, with average line widths of $\sim$1000 km s$^{-1}$ and $\sim$200 km s$^{-1}$ (FWHM). The measured line blueshifts are generally consistent with the outflow velocities of the absorption lines observed in the UV spectrum of NGC 4151 (Kriss \etal 1995), implying that the UV and X-ray line features originate in the same material.

\item Plasma diagnostics, based on the ratios of the resonance, intercombination and forbidden lines of the observed helium-like triplets, imply electron temperatures of $\sim$1-5$\times10^{4}$ K and electron densities of $\sim$10$^{8}$ cm$^{-3}$ in nitrogen and oxygen and $\sim$10$^{10}$ cm$^{-3}$ from neon (Porquet \& Dubau 2000) We note again that the effects of photoexcitation may have a considerable effect on the values of electron temperature and density determined by these diagnostic ratios (see Section~\ref{3-3}). The wide range of the calculated electron densities is suggestive of either a `clumpy' extended NLR or a region in which the material has grossly non-solar elemental abundances.

\item The soft X-ray spectrum of NGC 4151 is remarkably similar to the soft X-ray spectrum of NGC 1068 (Kinkhabwala \etal 2002). The correspondence between the two source spectra is evident on the general scale from the strong emission lines and RRC associated with the same ionic species, which dominate the spectra of both sources. The resemblance of the NGC 4151 and NGC 1068 soft X-ray spectra is also apparent in the details of the spectra ({\it i.e.} the similar helium-like triplet ratios and RRC temperatures).

\item The `ionization cone' model, developed by Kinkhabwala \etal (2002) to model the RGS spectra of NGC 1068, reproduces all of the hydrogen-like and helium-like emission features observed in the soft X-ray spectrum of NGC 4151 {\it in detail}. The best-fit radial and transverse velocities of the material in the ionization cones agree well with the blueshift and velocity dispersion measured from the individual line centroid energies, and are similar to the radial and transverse velocities measured from the soft X-ray spectrum of NGC 1068. The individual ionic column densities are somewhat smaller than those measured for NGC 1068, which, in combination with the plasma diagnostic ratios from the helium-like triplets, suggests that the material in the ionization cones of NGC 4151 may be somewhat denser than the material in the ionization cones of NGC 1068.

\end{itemize}

\section{Acknowledgments}

NJS gratefully acknowledges financial support from PPARC and, with REG, also acknowledges the support of NASA grant NAG5-9902. It is a pleasure to thank Keith Arnaud for his help with XSPEC. This work is based on observations obtained with XMM-Newton, an ESA science mission with instruments and contributions directly funded by ESA Member States and the USA (NASA). The authors wish to thank the \xmm ~team for the hard work and dedication that underlies the success of the mission. This research has made extensive use of NASA's Astrophysics Data System Abstract Service.

\end{document}